# Localizing the Origin of Idiopathic Ventricular Arrhythmia from ECG Using an Attention-Based Recurrent Convolutional Neural Network

*Mohammadreza Shahsavari, Niloufar Delfan,*
*and Mohamad Forouzanfar, Senior Member, IEEE*

**Abstract — Idiopathic ventricular arrhythmia (IVAs) are extra abnormal heartbeats disturbing the regular heart rhythm that can become fetal if left untreated. Cardiac catheter ablation is the standard approach to treat IVAs, however, a crucial prerequisite for the ablation is the localization of IVAs' origin. The current IVA localization techniques are invasive, rely on expert interpretation, or are inaccurate. In this study, we developed a new deep learning algorithm that can automatically identify the origin of IVAs from ECG signals without the need for expert manual analysis. Our developed deep learning algorithm was comprised of a spatial fusion to extract the most informative features from multichannel ECG data, temporal modeling to capture the evolving pattern of the ECG time series, and an attention mechanism to weigh the most important temporal features and improve model interpretability. The algorithm was validated on a 12-lead ECG dataset collected from 334 patients (230 female, age 46±13) who experienced IVA and successfully underwent a catheter ablation procedure that determined IVA's exact origins. The proposed method achieved an area under the curve of 93%, an accuracy of 94%, a sensitivity of 97%, a precision of 95%, and an F1-score of 96% in locating the origin of IVAs and outperformed existing automatic and semi-automatic algorithms. The proposed method shows promise toward automatic and noninvasive evaluation of IVA patients before cardiac catheter ablation.**

This work was supported in part by the by the Natural Sciences and Engineering Research Council of Canada (NSERC) under grant RGPIN-2021-03924 (to M.F.)

M. Shahsavari and N. Delfan are with the Department of Systems Engineering, École de technologie supérieure (ÉTS), Université du Québec, Montréal, QC, Canada.

M. Forouzanfar is with the Department of Systems Engineering, École de technologie supérieure (ÉTS), Université du Québec, Montréal, QC, Canada, and Centre de recherche de l'Institut universitaire de gériatrie de Montréal (CRIUGM), Montréal, Canada (e-mail: mohamad.forouzanfar@etsmtl.ca).

## I. Introduction

Premature ventricular complex (PVC) is defined as a premature heartbeat occurring when the lower chambers of the heart contract too early [1]. Ventricular tachycardia (VT) is a heart rhythm disorder (arrhythmia) defined as three or more PVCs in a row, at a rate of more than 100 PVC beats per minute. For normal individuals, an occasional period of PVC is not considered a problem and usually does not need treatment [2]. However, PVCs become more of a concern, if they happen frequently or when other heart problems are present. For example, for an individual whose ventricle already squeezes poorly, PVCs may be life-threatening. Idiopathic ventricular arrhythmias (IVAs) are those PVCs and VTs which occur for an unknown reason and in the absence of structural heart disease.

Cardiac catheter ablation is usually performed to prevent IVAs and to restore a normal heart rhythm [3]. This procedure consists of two main stages, a diagnostic stage in which the origins of electrical signals causing IVAs is be found, and a treatment stage where the abnormal tissue is ablated (destroyed) through heating or freezing. Thus, it is very important to correctly locate the origin of the abnormal heart electrical activity before ablation. The origin of these abnormal signals is mainly located in either the right ventricular outflow tract (RVOT) or the left ventricular outflow tract (LVOT) [2, 4].

The conventional methods used for the detection of IVA origin include electrical mapping, substrate mapping, and pace mapping. Electrical mapping [5, 6] is performed by moving an electrically sensitive catheter to different ventricle points and recording electrical signals inside the heart (electrograms). The recorded electrograms are analyzed to identify the IVA electrical pathway. This technique can only be performed in a small number of patients who can withstand a stable IVA for the entire mapping duration. Substrate mapping [7-11] studies the electrical properties of the ventricle to identify the critical and slow conduction areas. This is performed by detecting abnormal electrograms obtained in normal sinus rhythm and/or after stimulation of the ventricles from various sites [8, 9]. However, the presence of such abnormal electrograms at a specific site does not necessarily mean that this site is the IVA origin [12]. Besides, the abnormal electrograms caused by IVAs are low-amplitude and are hard to distinguish from noise in dense fibrotic scar areas. Pace mapping is a substitute technique for electrical mapping [12-15] performed in two steps. First, a clinical ECG record is obtained when an IVA happens. Next, a catheter is used for stimulating the heart from different ventricular sites and to produce electrical pathways originating from these sites. The IVA's origin is recognized when the electrical pathway generated by stimulation best matches that of the clinical IVA. Compared to electrical mapping, this method gives a better sense of IVA origin by reconstructing and visualizing the actual IVA electrical pathway. The conventional methods for the detection of IVA are invasive, high risk to the patient, complicated, time-consuming, and expensive.

Recently, new non-invasive methods for the detection of IVA origins have been proposed based on the analysis of ECG signals. Several studies have already proven a strong relationship between the features of electrocardiogram (ECG) and the locations where IVAs stem from [16-19], and several approaches have been developed to find the IVAs' origins based on this relationship [20-24]. For example, in [20], a threshold based on the amplitude of ECG R wave in lead I was used to identify the IVA origin. In [21], a mathematical model based on S wave and R wave amplitudes was introduced as an index for the identification of the IVA origin. In [22], ECG QRS morphological features were extracted by three electrophysiologists, and an extreme gradient boosting tree classifier was designed for the detection of IVA origins. In [23], a 12-lead ECG dataset was simulated using a computer heart model to train a convolutional neural network for the localization of IVA. In [24] a support vector machine (SVM) was designed to locate IVA origins. The use of convolutional neural networks was also investigated in [24], however, the designed network could not outperform the conventional SVM model.

Given that automatic detection of morphological features in an abnormal ECG is challenging, most IVA localization methods rely on manual ECG morphological feature extraction and analysis [17, 20, 21, 25, 26]. Among automatic IVA localization algorithms, some rely on simulated data that may not fully represent the real ECG [23] while others rely on conventional machine learning algorithms that are not capable of fully extracting the ECG spatial and temporal information [24]. There is therefore a significant demand for a fully automatic approach that can detect the origin of IVA from the complex spatial-temporal pattern of ECG multichannel data without the need for any manual analysis. Among different automatic techniques for the characterizing of ECG multichannel data, deep learning has shown greater potential and more accurate results in various applications [27, 28].

In this paper, we propose an end-to-end deep learning-based framework trained on a real-world dataset to automatically localize IVA origins. Our feature extraction and training phase are end-to-end which gives the algorithm the ability to automatically extract the most informative features correlated with IVA origins. Our deep learning framework relies on a combination of convolutional neural networks (CNNs) and recurrent neural networks (RNNs) to capture the spatial information contained within the ECG multichannel data and the temporal information contained in the ECG time pattern. An attention mechanism is employed to further focus the algorithm on the most important segments of ECG signal and improve model interpretability. The proposed framework is validated on a dataset of 334 patients using a 10-fold cross-validation approach to ensure the reliability and generalizability of the reported results.

## II. METHOD

### A. Dataset

This study was conducted on a publicly available 12-lead ECG dataset collected from 334 patients who experienced IVA and successfully underwent a catheter ablation procedure to validate the exact origin of IVAs [2].

The dataset has been collected under the auspices of Chapman University and Ningbo First Hospital of Zhejiang University. The institutional review board of Ningbo First Hospital of Zhejiang University has approved this study and has allowed the data to be shared publicly after de-identification.

The average length of ECG recordings in the dataset is $10.53 \pm 3.58$ s and the sampling rate is 2000 Hz. Of the 334 patients, 77% are classified as RVOT and 23% as LVOT. Participants' characteristics are provided in Table I.

### B. Problem Formulation

Finding the origin of IVA can be formulated as a time-series classification problem in which a comprehensive model is required to extract useful information from varied-length ECG records and predict the correct class for each record. We transformed this problem into a deep learning framework that receives a varied-length ECG as input, and outputs a single binary value, indicating LVOT or RVOT. The objective of the deep learning framework was to minimize the binary cross-entropy between the reference labels and models outputs, given by:

$$Loss = -\frac{1}{N}\sum_{i=1}^{N}(y_i . \log y_i^{'} + (1-y_i).\log(1-y_i^{'}))$$

where $y_i$ is the reference label, $y_i^{'}$ is the model output for the $i$ th ECG record, and $N$ is the total number of available ECG segments. Binary cross-entropy loss was chosen to minimize the distance between two predicted and actual probability distributions.

TABLE I
PATIENTS CHARACTERISTICS

| Characteristics | All | RVOT | LVOT |
|---|---|---|---|
| Patients, n (%) | 334 | 257 (77) | 77 (23) |
| Age, Mean ± SD, year | 46.1 ± 13.1 | 47.5 ± 13.4 | 46.2 ± 16.5 |
| Male, n (%) | 104 (32) | 65 (27) | 39 (49) |
| Frequent PVC, n (%) | 325 (98) | 251 (99) | 74 (97) |
| Sustained VT, n (%) | 9 (2) | 6 (1) | 3(3) |

## C. Preprocessing

Our dataset contained ECG signals with different lengths (5 s-25 s). To have a fixed input size, all ECG signals were zero-padded to the length of the longest ECG signal (25 s). Padding ECG signals allows the application of deep learning algorithms that require input samples with equal lengths. ECG signals were then lowpass filtered at 25 Hz with a 4th order Butterworth filter to remove high-frequency noise and downsampled to 50 Hz to reduce the input data dimension. The filter was applied in forward and backward directions to achieve a zero-phase shift. ECG segments were normalized by removing the mean and scaling to unit variance. As ECG data can have a wide range of values, normalization eases the learning process of the deep learning algorithm [29].

## D. Model Architecture

Deep learning has proved its significant potential in predictive modeling of complex physiological processes including different cardiovascular signals such as ECG [30-32]. Here, we designed a novel deep neural network (DNN) structure specific to multichannel ECG data by combining CNN, LSTM, and MLP architectures. It involved four different processing and learning stages including spatial fusion, temporal modeling, attention mechanism, and fully connected layers which enabled the optimum extraction of spatial and temporal features from the ECG multichannel data and the automatic modeling of their relationship with the IVA origin (see Fig. 1).

***Spatial Fusion:*** To extract the relevant spatial information contained in the 12 leads of the ECG signal, a 13-layer CNN based on the VGG structure [33] was designed. VGG has proved as a powerful spatial feature extractor even for 1-D signals [30]. The designed network encoded the 12 ECG channels of size 1250 into 512 channels of size 78. The encoded data contained important spatial information that were further processed for IVA localization.

***Temporal Modeling:*** To capture the temporal information and to learn the complex dependencies in the ECG time sequence, two bidirectional long short-term memory (BiLSTM) [34, 35] were designed. Each BiLSTM layer consisted of 2 LSTM layers: one that iterated the input time series forward in time and another one that iterated them backward. The BiLSTM layers consisted of 64 LSTM units that received 78 encoded samples from the spatial fusion output one by one in 78-time steps. In each time step, the temporal model read all the 512 channels. After processing each time step, it returned two vectors of size 64 called BiLSTM states containing the extracted temporal information from the sequence until that time step, one corresponding to the forward LSTM layer and the other to the backward LSTM layer.

***Attention Mechanism:*** The attention mechanism used in this work was based on a similar principle as the one used in [36]. Rather than focusing on BiLSTM final state, attention calculated a weighted average of BiLSTM states in all time steps. This was performed by using two fully connected layers along with a hyperbolic tangent activation function that were applied to BiLSTM states in all time steps. These fully connected layers learned to assign a score to each time step in the input time series. A Softmax operation was then applied to

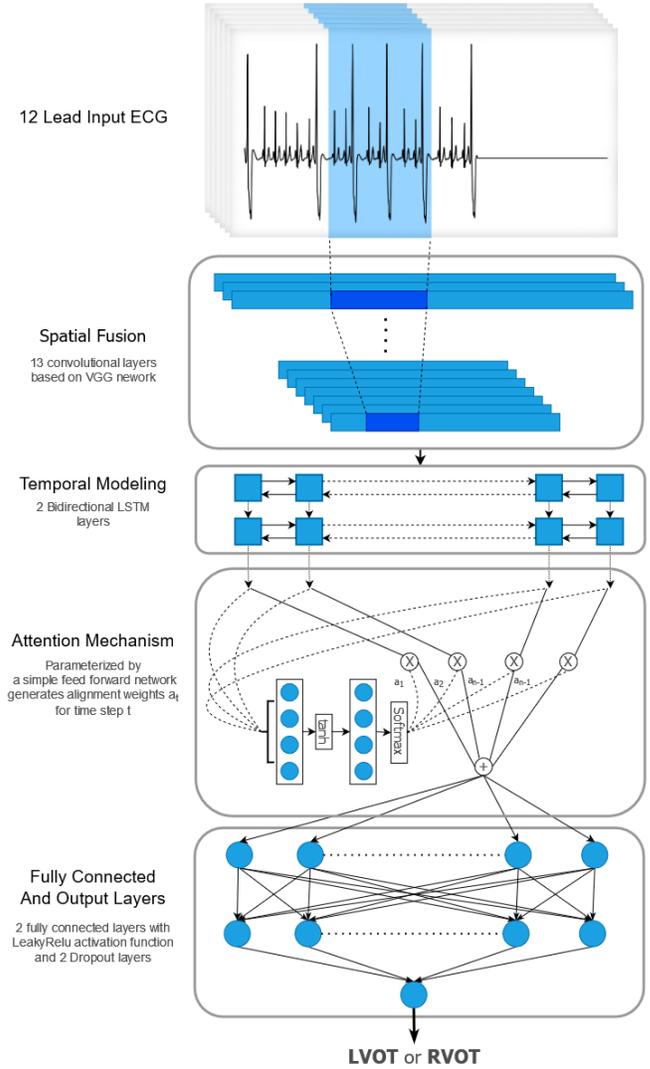

**12 Lead Input ECG**

**Spatial Fusion**
13 convolutional layers
based on VGG network

**Temporal Modeling**
2 Bidirectional LSTM
layers

**Attention Mechanism**
Parameterized by
a simple feed forward network
generates alignment weights $a_t$
for time step t

**Fully Connected
And Output Layers**
2 fully connected layers with
LeakyRelu activation function
and 2 Dropout layers

**LVOT** or **RVOT**

Fig. 1. Our proposed DNN model for localizing the IVA origin.

these scores to designate the importance of each time step and generate attention weights. The designated weights focused the model's attention on the most informative parts of the ECG signal and therefore improved performance and enhanced interpretability.

***Fully Connected and Output Layers:*** The attention output was passed to two fully connected layers with 256 nodes and LeakyReLU activation functions with a 0.01 negative slope. The fully connected layer was responsible for combining all the features extracted in the previous layers and preparing a more abstract feature representation needed to make the final decision in the output layer. Two dropout layers with the value of 0.2 were used to prevent the algorithm from overfitting.

An output layer with sigmoid activation functions, to empower the network learning the complex and nonlinear relationship between the inputs and the targets, was used to classify LVOT from RVOT.

## TABLE II
### DNN Model Hyperparameters

| Hyperparameter | Search interval | Optimum value |
|---|---|---|
| Number of convolution layers | 3 to 15 | 13 |
| Number of BiLSTM layers | 1 to 3 | 2 |
| Number of LSTM units | 32 to 512 | 64 |
| Number of nodes in the attention layer | 16 to 128 | 64 |
| Number of fully connected layers | 1 to 3 | 2 |
| Number of nodes in the fully connected layer | 128 to 1024 | 256 |
| Learning rate | 0.00001 to 0.01 | 0.00005 |
| Batch size | 32 or 64 | 32 |

### *E. Data Analysis*

A 10-fold cross-validation was performed to reliably evaluate the generalizability performance of the proposed method on unseen data. In each fold, 80% of data were used for training, 10% of data were used as a validation set to find the model optimum hyperparameters, and 10% were used to evaluate the performance of the optimum trained network on unseen data. This procedure was then repeated 10 times so that data belonging to every individual was once placed in the test set. The Adam optimization algorithm was used to minimize the binary cross-entropy loss function.

A random search was performed to find the optimum hyperparameters of our model. The model was trained using different combinations of hyperparameter values and the set with the best validation performance was selected. The list of model hyperparameters is listed in Table II.

## III. Results

We evaluated our proposed algorithm for localization of the origin of IVAs on data collected from 334 individuals (see Table I) using a 10-fold cross-validation.

The optimum model hyperparameters selected based on the validation results are listed in Table II. Our experiments revealed a DNN structure with 13 convolution layers, 2 BiLSTM layers with 64 LSTM units, an attention layer with 64 nodes, and 2 fully connected layers with 256 nodes and LeakyReLU activation functions with negative slopes of 0.01 leads to the best generalization performance. The optimum training parameters included a batch size of 32 and a learning rate of 0.00005.

The achieved results on the test sets (not seen during training and validation) are reported in Table III. The results are reported for every fold as well as the whole dataset in terms of the area under the curve (AUC), accuracy (ACC), sensitivity (SE) or recall, specificity (SP), and F1-score (F1). It is observed that depending on the train/validation/test split the localization results vary. For example, the accuracy varied from 84.8% in fold 7 to 100% in fold 3 while the average

## TABLE III
### Performance of Our Proposed DNN Framework for the Identification of IVA Origin. The Results are Reported on the Test Set not Seen During the Training and Validation Phase.

| Fold Number | AUC | ACC | SE | SP | PPV | NPV | F1 |
|---|---|---|---|---|---|---|---|
| 1 | 0.972 | 0.969 | 1 | 0.857 | 0.962 | 1 | 0.981 |
| 2 | 0.982 | 0.969 | 0.965 | 1 | 1 | 0.8 | 0.982 |
| 3 | 1 | 1 | 1 | 1 | 1 | 1 | 1 |
| 4 | 0.965 | 0.909 | 0.956 | 0.8 | 0.916 | 0.888 | 0.9361 |
| 5 | 1 | 1 | 1 | 1 | 1 | 1 | 1 |
| 6 | 0.944 | 0.939 | 1 | 0.666 | 0.931 | 1 | 0.964 |
| 7 | 0.815 | 0.848 | 0.84 | 0.875 | 0.954 | 0.636 | 0.893 |
| 8 | 0.827 | 0.939 | 1 | 0.666 | 0.931 | 1 | 0.964 |
| 9 | 0.951 | 0.945 | 1 | 0.818 | 0.928 | 1 | 0.962 |
| 10 | 0.847 | 0.909 | 0.954 | 0.818 | 0.913 | 0.9 | 0.933 |
| **Total** | **0.933** | **0.943** | **0.972** | **0.844** | **0.954** | **0.902** | **0.963** |

## TABLE IV
### Comparisons of Our Proposed Deep Learning Approach with State-of-the-art IVA Localization Techniques. All the Methods Are Implemented on the Same Dataset Using Same Train, Validation and Test Strategy

| Author | Method | Performance | | | | | |
|---|---|---|---|---|---|---|---|
| | | AUC | ACC | SE | SP | PPV | F1 |
| Yang et al. (2017) [23] | Automatic spatial feature extraction using CNN | 0.74 | 0.80 | 0.94 | 0.34 | 0.83 | 0.88 |
| Nakam-ura al (2021) [24] | Automatic spatial feature extraction using CNN | 0.68 | 0.81 | **0.97** | 0.26 | 0.81 | 0.88 |
| | SVM classifier | 0.57 | 0.71 | 0.84 | 0.31 | 0.80 | 0.82 |
| This study | Attention-based automatic spatial and temporal feature extraction using CNN-BiLSTM | **0.93** | **0.94** | **0.97** | **0.84** | **0.95** | **0.96** |

accuracy over all the folds was 94.3%. A similar pattern is also exhibited for other classification metrics. Overall, the proposed method achieved relatively high performance in terms of all the evaluated metrics.

Our proposed method was compared with three state-of-the-art techniques on the same dataset using the same train, validation, and test sets. The results are reported in Table IV. It is observed that the proposed method substantially outperforms the other techniques. The achieved improvements can be attributed to our adopted hybrid deep architecture that extracted both within and between channel information from ECG data, and the attention mechanism that further focused the model's emphasis on important features.

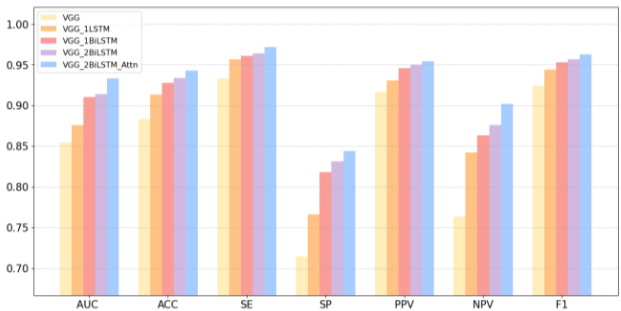

Fig. 2. Performance of different modules of our DNN framework on its overall performance.

An ablation study was also performed to evaluate the performance of different modules of our DNN framework on its overall performance. Fig. 2 compares the performance of our DNN framework when using only the VGG model, VGG and one LSTM layer, VGG and one BiLSTM layer, VGG and two BiLSTM layers, and VGG and two BiLSTM layers with the attention mechanism. The AUC, ACC, SE, SP, PPV, NPV, and F1 were 85%, 88%, 93%, 71%, 92%, 76%, and 92% using the VGG network, 88%, 91%, 96%, 77%, 93%, 84%, and 94% by adding a one-layer LSTM, 91%, 93%, 96%, 82%, 95%, 86%, 95%, by adding a one-layer BiLSTM, 91%, 93%, 96%, 83%, 95%, 88%, and 96% by adding a two-layer BiLSTM, and 93%, 94%, 97%, 84%, 95%, 90%, and 96% by adding an attention mechanism. It is observed that by adding the recurrent architectures the performance is improved and the best results are achieved when combining the VGG network with two BiLSTM layers and the attention mechanism. For example, the AUC was improved by 6% after adding the recurrent layers and by another 2% after adding the attention mechanism.

## IV. DISCUSSION

Twelve-lead ECG is considered the preferred technique for identifying abnormal cardiac conditions [37] as it can be easily performed by placing 10 surface electrodes on the patient's limbs and chest. The use of ECG as a noninvasive tool in the assessment of ventricular arrhythmias is of particular importance for diagnosis and treatment planning. However, the analysis of ECG signals requires highly trained experts.

In this study, we investigated the application of deep learning to identify the origin of IVA from 12-lead ECG data without the need for expert interpretation. A DNN comprising several learning modules including a CNN architecture for spatial feature extraction, an RNN for modeling the time pattern evolution, and an attention mechanism to weigh the most important ECG segments was designed. It was shown that the developed approach can effectively unveil the relationship between the spatiotemporal pattern of multichannel ECG data and the origin of IVA with an AUC of 93.3%. To evaluate the performance of our methods, we used several classification metrics including AUC, ACC, SE, SP,

PPV, NPV, and F1-Score. AUC shows the separability of LVOT and RVOT classes. An AUC of closer to 1 means better classification performance on both RVOT and LVOT prediction. ACC measures the proportion of correctly classified cases. While it is simple and easy to understand, it cannot give a good understanding of the model performance, especially on a balanced dataset. SE shows how well the model can classify RVOT cases correctly while SP indicates the model's ability to correctly classify LVOT cases. PPV predicts how likely someone is to be classified as RVOT, while NPV predicts how likely someone is to be classified as LVOT. F1-Score is the harmonic mean of sensitivity and PPV which is especially useful when data is imbalanced. Given that our dataset was imbalanced (257 RVOT cases and 77 LVOT cases) AUC and F1-score can provide a better demonstration of our model performance. In addition, given that the detection of LVOT and RVOT are equally important, SE and SP are of equally high importance.

An ablation study revealed an improved averaged performance of about 3.5% (average of all evaluation metrics) when adding an RNN (LSTM) module to the CNN (VGG) network. The results were further improved by about 2.8% when using a more powerful two-layer BiLSTM network (see Fig. 2). Adding an attention mechanism further improved the performance by approximately 1.2%, on average. These results show that a hybrid model that can effectively capture the spatiotemporal information hidden in the pattern of multichannel ECG data performs the best among different deep architectures for localizing the origin of IVAs.

Table V compares the proposed approach with other conventional and state-of-the-art techniques for localizing the origin of IVA. Unlike conventional clinical approaches such as electrical mapping [5, 6], substrate mapping [7-11], and pace mapping [12-14] that are invasive, our proposed method is solely based on the analysis of noninvasively measured ECG and therefore poses no significant risk to the patient. Unlike manual approaches that rely on the manual extraction and analysis of ECG morphological features such as R and S wave amplitudes and timings [17, 20, 21, 26], our proposed method automatically learns from the spatiotemporal pattern of ECG multichannel data and therefore does not require expert manual analysis. Unlike semi-automatic machine learning techniques that learn from manually extracted ECG features [23, 25], our algorithm automatically extracts and learns ECG features and therefore is less expensive and time-consuming. Among fully automatic techniques, our algorithm is the only algorithm that is validated on real data (unlike [23] which used simulated data), extracts both spatial and temporal features of the ECG signal (unlike [24] that only modeled the ECG spatial information) and is fully validated on all available data using a 10-fold cross-validation. The closest results among the existing automatic algorithms are reported in [24] on a dataset of 464 individuals using an SVM model without specific feature extraction. The same approach was implemented and tested on our dataset (see Table IV). Given that the reported results in [24] were calculated on a single hold-out test set, we conjecture that they may not be fully generalizable.

TABLE V
COMPARISONS OF DIFFERENT ECG-BASED ANALYSIS METHODS FOR LOCALIZING OF THE ORIGIN OF IVA

| Category | Author | Method | Limitations |
|---|---|---|---|
| Manual | Cheng et al. (2018) [17] | Defined an index based on R-wave and S-wave amplitudes in right precordial and posterior ECG leads | Relied on experts to identify and quantify R and S waves |
| | Di et al. (2019)[21] | Defined an index based on R-wave and S-wave amplitudes in leads $V_1$, $V_2$ and $V_3$ | Relied on experts to identify and quantify R and S waves |
| | Wang et al. (2021) [26] | Defined an index based on R-wave amplitude in lead I | Relied on experts to identify and measure R wave |
| Semi-automatic | He et al. (2018) [25] | Logistic regression algorithm using R-wave and S-wave features | -Relied on experts to identify and quantify R and S waves<br>-Single hold-out evaluation |
| | Zheng, Fu et al. (2021) [22] | Extreme gradient boosting tree based on 1,600,800 features extracted from ECG. | -QRS complex was manually extracted before applying machine learning.<br>-Single hold-out evaluation |
| Automatic | Yang, Yu et al. (2017) [23] | Automatic spatial feature extraction using CNN | -Validated on simulated data (17090)<br>-Only relied on the extraction of spatial features<br>-A separate validation set was not used to adjust model hyperparameters |
| | Nakamura, Nagata et al (2021) [24] | Automatic spatial feature extraction using CNN | -Only relied on the extraction of spatial features<br>-Single hold-out evaluation<br>-Limited number of patients |
| | | SVM classifier | -Single hold-out evaluation<br>-Limited number of patients (464) |
| | This study | Attention-based automatic spatial and temporal feature extraction using CNN-BiLSTM | -Limited number of patients (334) |

*Results are reported on our own dataset.

Our method was tested on a modest cohort of 334 individuals with IVA. As the available data was somewhat limited, a cross-validation technique was used to evaluate the generalizability of the model when encountering unseen data. The significance of our cross-validation approach can be observed in Table III, where it is shown that depending on the selection of the test set, the detection performance can vary. The best-achieved performance (AUC) was up to 100% while by changing the distribution of train and test data the performance was decreased to nearly 80% in some folds. To ensure a fair and unbiased evaluation, we reported the average performance among different folds. This approach is superior to the previous validation methods that held out a specific small portion of data for testing the performance [24]. Using a single hold-out set as the test data can lead to an unreliable (and unstable) performance evaluation.

We used an attention mechanism to 1) potentially improve localization performance and 2) enhance the interpretability of the algorithm. It was observed that by adding attention to the network, AUC, ACC, SE, SP, PPV, NPV, and F1 are improved by 1.9%, 0.9%, 0.8%, 1.3%, 0.4%, 2.6%, and 0.6%. The assigned weights of the attention mechanism to different parts of the ECG signal over time can illustrate their importance in localizing the origin of IVAs. Fig. 3 shows examples of 12-lead ECGs for individuals with RVOT and LVOT. The attention weights are plotted in red on top of the figures. It is observed that attention is focusing on specific parts of the input data that are mostly related to abnormal heartbeats. On the other hand, some irrelevant parts of the ECG signal such as the zero-padded segments are completely ignored by the attention mechanism.

A limitation of the current study was the relatively limited number of individuals (334) used for performance evaluation. Our dataset was also imbalanced in terms of the target classes (257 patients with RVOT out of 334) and gender distribution (230 females out of 334 individuals). To generate more training data, we tried to augment ECG signals by shifting or squeezing and stretching them along the time axis. However, no improvement in the localization of IVAs was achieved. Future work should focus on more advanced augmentation techniques as well as generating synthetic ECG signals using generative adversarial networks (GANs) [38]. Generating realistic synthetic ECG signals not only helps to provide a larger dataset but can also make the dataset balanced with respect to the target classes or even genders. A semi-supervised approach is also recommended to take advantage of both labeled and unlabeled data [39]. There exist several public datasets containing ECGs with PVC or VT but their origins are not labeled [37]. These unlabeled data can be used in a semi-supervised learning framework along with limited labeled data to achieve a higher localization performance.

The proposed method was solely tested on localization of right and left IVAs. As a future work, a further detailed set of data with precise location of IVA should be collected to evaluate the performance of the proposed method in localization of the exact IVA origins.

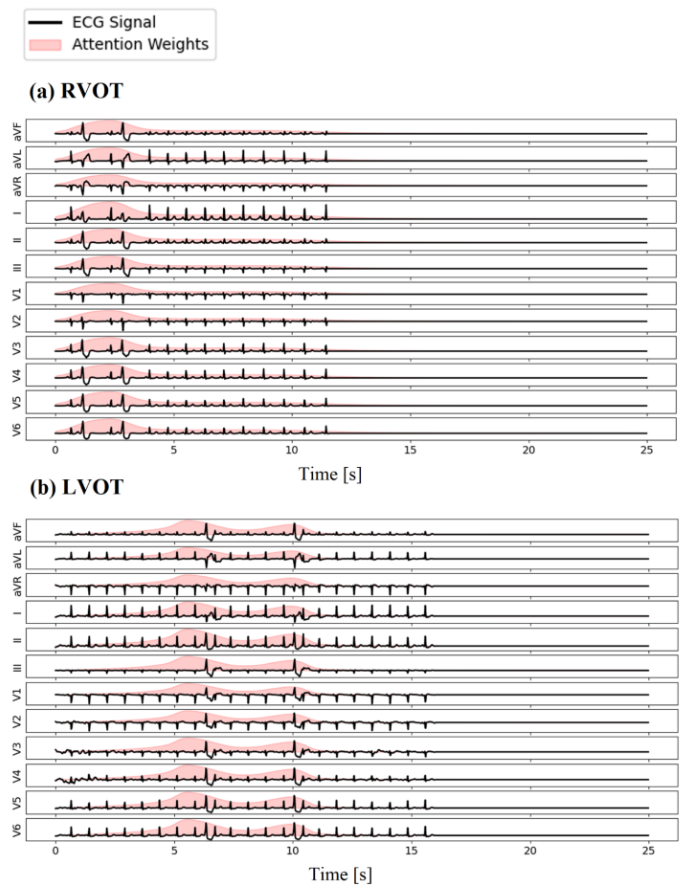

Fig. 3. Examples of 12-lead ECG signals for patients with IVAs located at a) right ventricular outflow tract (RVOT) and b) left ventricular outflow tract (LVOT). The attention weights are highlighted in red on top of the ECG signals.

## V. Conclusion

We proposed a fully automatic deep learning model to identify IVA origin from noninvasively measured 12-lead ECG. Our algorithm achieved a high localization performance (AUC = 93.3%) and outperformed existing techniques. Our proposed algorithm was based on spatial modeling using a VGG CNN, temporal modeling using a two-layer BiLSTM RNN, and temporal weighting using an attention mechanism. Unlike manual and semi-automatic algorithms, the proposed approach provided end-to-end automatic processing of ECG data without the need for any expert analysis. Given the short-duration ECG signals (~10 sec) required for the analysis, our deep learning model provides a cost-effective and low-risk approach toward the identification of the origin of IVAs before cardiac catheter ablation.